\documentstyle{article}

\newcommand{\beq}{\begin{equation}}
\newcommand{\eeq}{\end{equation}}

\title{ \bf Replica Fourier Transforms on ultrametric trees, \\
        and block-diagonalizing multi-replica matrices. }

\author{
\\
  { C.De Dominicis}              \\
  {\small\it S.Ph.T., CE Saclay } \\[-0.2cm]
  {\small\it 1991 Gif sur Yvette , France}          \\
  {\small Internet: {\tt CIRANO@SPHT.SACLAY.CEA.FR}}     \\
 \\  
  {D. M. Carlucci}               \\
  {\small\it Scuola Normale Superiore di Pisa}  \\[-0.2cm]
  {\small\it Piazza dei Cavalieri}        \\[-0.2cm]
  {\small\it Pisa 56126, Italy}          \\
  {\small Internet: {\tt CARLUCCI@UX2SNS.SNS.IT}}     \\ 
 \\        
  { T. Temesv\'ari}              \\
  {\small\it Institute for Theoretical Physics} \\[-0.2cm]
  {\small\it E\"otv\"os University } \\[-0.2cm]
  {\small\it 1088 Budapest, Hungary}          \\
  {\small Internet: {\tt TEMTAM@HAL9000.ELTE.HU}}     \\
       }

\date{\today}

\begin{document}

\maketitle

\abstract{The analysis of objects living on ultrametric trees, in particular the 
 block-diagonalization of $4-$replica matrices $M^{\alpha \beta ; \gamma \delta}$, 
 is shown to be dramatically 
 simplified through the introduction of properly chosen operations on those 
 objects. These are the Replica Fourier Transforms on ultrametric trees. 
 Those transformations are defined and used in the present work.}

\clearpage

\begin{center} 
    {\bf Resum\'e} 
\end{center} 

On montre que l'analyse d'objets vivant sur un arbre ultrametrique, en 
particulier, la diagonalisation par blocs d'une matrice 
$M^{\alpha \beta ; \gamma \delta}$ dependant de $4-$repliques, se simplifie 
de fa\c con dramatique si l'on introduit les operation appropri\'ees sur ces 
objects. Ce sont les Transform\'ee de Fourier de Repliques sur un arbre 
ultrametrique. Ces transformations sont definies et utilis\'ees dans le 
present travail.

\clearpage

Spin glasses and their typical glassy phases appear to be present in 
a wide spectrum of domains\mbox{\cite{Mezard1}-\cite{Halpin}}. 
The high level of complexity  
inherent to their structure (field $\phi_{\alpha \beta}(\vec{x})$ 
depending upon two replicas, propagators 
$G^{\alpha \beta; \gamma \delta}(\vec{x},\vec{y})$ depending upon four 
of them) has prevented so far a systematic study of the glassy phase 
with the standard tools of field theory and in particular the 
renormalization group where objects with $6$ (or $8$) replicas would be 
needed. One recent step in the process of 
founding a field theory has been the spelling out of a Dyson like 
equation\cite{Cyrano1, Temesvari} relating the propagators $G$ 
and their inverses $M$
(mass operator plus kinetic terms), a triviality in standard field 
theories. To be more precise, the obtained relationship is not directly 
between $G$ and $M$, but between what was termed the kernel ($F$) of 
$G$ and the kernel ($K$) of $M$. What makes the algebra complicated, 
is that one cannot directly work with replicas (whose number $n$ is to be set 
to $0$). Instead, one is led to use, for the observables, a replica 
symmetry broken representation (RSB) usually inferred
from mean field studies (be it with $R=1$, one step RSB or $R=\infty $ 
as in Parisi's\cite{Parisi1} ansatz). As a result, 
the inversion process of the 
$4-$replica matrix $M^{\alpha \beta ; \gamma \delta}$, becomes a 
highly non trivial exercise with results fairly involved 
for $R=\infty$\cite{Cyrano1}, and even much more so for a generic 
$R$\cite{Temesvari}.

Here we would like to show that with the appropriate choice of 
transformations, the previously obtained results, considerably simplify, 
and in some sense, become almost transparent.

To that effect we introduce the notion of $(i)$ Replica Fourier 
Transform (RFT) and $(ii)$ RFT on a (ultrametric) tree.  
With those definitions, given in section 1, we show that the 
relationship between (matrix) functions and their kernels is 
nothing but a Replica Fourier Transformation (a double transform for the 
Replicon component and a single one for the longitudinal 
anomalous component). This is done in section 2 and 3. 
Further the same RFT, used now on the eigenvalue equations immediately 
(block-)diagonalizes them (section 4). 
Used on the equation $GM=1$, the RFT yields Dyson's equation 
relating the corresponding kernels of $M$ and $G$ (section 5).

It must be mentioned at the outset that this work was prompted
by the yet unpublished results of Parisi and Sourlas\cite{Parisi2} 
where a Fourier transform on $p-$adic numbers is introduced\footnote{
See also B.Grossman, ref \cite{Grossman}.}, which
diagonalizes the Replicon sector components. The RFT used here
has the flexibility that also allows for a block diagonalization of the 
Longitudinal Anomalous sector.

\section{Replica Fourier Transform}

It is assumed, as in Parisi\cite{Parisi1}, that the $2-$replica object 
$\langle \phi_{\alpha \beta}(\vec{x}) \rangle \equiv q_{\alpha \beta} 
= q_t$ only depends upon the overlap (or codistance  
$\alpha \cap \beta = t$). Likewise a $m-$replica object will 
depend upon $m-1$ independent overlaps. 
The RFT\cite{Carlucci}, 
which is a discretized version of the algebra introduced 
by Mezard and Parisi\cite{Mezard2} defines the RFT transform $\widehat{A}$ of 
$A$ via

    \beq
       \widehat{A}_k = \sum_{t=k}^{R+1} p_t 
       \left( 
             A_t - A_{t-1}
       \right) 
    \eeq
Here the $p_t$'s are the sizes of the Parisi boxes, $p_0\equiv n$ the 
replica number, $p_{R+1} \equiv 1$. Note that $\alpha \cap \beta =t $ 
means that $\alpha$ and $\beta$ belong to the same Parisi box of size $p_t$, 
but to two distinct boxes of size $p_{t+1}$, {\it i.e.} they belong to the 
(relative) multiplicity 

    \beq
     \delta_t=p_t - p_{t+1}
    \eeq

    \beq 
     \delta_{R+1}=p_{R+1}
     \eeq

Objects with indices  out of the range of definition (here $(0,R+1)$) 
are taken as null. Conversely one has

    \beq
       A_j=\sum_{k=0}^j \frac{1}{p_k} 
       \left(
       	     \widehat{A}_k - \widehat{A}_{k+1}
       \right)
       \label{inverse_FT}
    \eeq
and the associated (relative) multiplicity in what will be termed  below 
``resolution'' (or pseudo-momentum) space,

    \begin{eqnarray} 
         \bar{\delta}_{k} &=& \frac{1}{p_k}-\frac{1}{p_{k-1}} \\
         \bar{\delta}_{0} &=& \frac{1}{p_0}
		\label{mult}
    \end{eqnarray} 

Characteristically, the convolution  

    \beq 
       \sum_{\gamma=1}^n A_{\alpha \gamma} B_{\gamma \beta} 
       =
       C_{\alpha \beta} 
    \eeq
becomes, after RFT 
    \beq
      \widehat{A}_k \widehat{B}_k=\widehat{C}_k
      \label{convolution}
    \eeq
If we take 
 
    \beq
       C_{\alpha \beta}=\delta_{\alpha \beta} 
    \eeq
we have 

   \begin{eqnarray} 
      \widehat{C}_k &=& 1 \\
      \widehat{A}_k &=& 1/\widehat{B}_k
   \end{eqnarray} 
This convolution property conveniently allows to write for example, 

   \beq
      \mbox{tr}(A)^s=n\sum_{k=0}^{R+1} \bar{\delta}_k 
      \left( \widehat{A}_k \right)^s,  
   \eeq
to be compared with 

   \beq 
      \sum_{\alpha \beta} \left(A_{\alpha \beta}\right)^s
      =
      n \sum_{j=0}^{R+1} \delta_j \left(A\right)_j^s .
   \eeq
One is often using a multiplicity $\mu(k)$, related to the (relative) 
multiplicity $\bar{\delta}_{k}$ introduced here by 

   \begin{eqnarray} 
        \mu(k) & = & n \bar{\delta}_k \\
        \mu(0) & = & n \frac{1}{p_0}=1 
   \end{eqnarray}

\section{RFT on a tree}

Despite their usefulness\cite{Carlucci}, the above introduced 
objects are not yet 
tailored to fit the complexity of situations arising in the study of 
functions of more than two replicas. We need to extend the RFT definition 
to the case where the two replicas ({\it i.e.} the overlap) summed over, move 
on a (ultrametric) tree in the presence of other {\it passive} overlaps. The 
simplest example is the $3-$replica function (see below section 4)

	 \beq
            f^{\alpha \beta; \mu} 
	    \equiv
	    f^{r}_t 
	 \eeq
where we have a (fixed) overlap $r = \alpha \cap \beta$, and the 
{\it lower index}, $t$ is the {\it cross-overlap} $t$

     \begin{eqnarray} 
        \max(\alpha \cap \mu ; \beta \cap \mu) & = & t \nonumber \\
        \alpha \cap \beta & = & r    
     \end{eqnarray} 
As we move $\mu$ {\it i.e.} $t$ along the $3-$replica tree, we now 
have the successive multiplicities

	 \begin{eqnarray} 
            p_t - p_{t+1}  &\,\, & t<r \nonumber \\
             p_t - 2p_{t+1}  &\,\, & t=r \\
            2(p_t - p_{t+1})  &\,\, & t>r \nonumber 
         \end{eqnarray}
Indeed, in the case $t=r$, there are two boxes of size $r+1$  excluded 
for $\mu$  (the box $p_{r+1}$ of $\alpha$, and the one of $\beta$). 
In the case $t>r$ there are two branches of the tree available for $\mu$ 
($\alpha \cap \mu=t$, $\beta \cap \mu=r $ or 
 $\alpha \cap \mu=r$, $\beta \cap \mu=t $). It is thus appropriate to define 
effective boxes in the presence of a passive, direct overlap $r$, {\it viz.} 
$p_t^{(r)}$ with

	  \begin{eqnarray} 
	    p_t^{(r)} = p_t & \, \, & t \leq r \nonumber \\ 
            p_t^{(r)} = 2p_t & \, \, & r < t
	    \label{poids_bis}
          \end{eqnarray} 
and the associated {\it RFT on the 3-replica tree}

    \beq 
       \widehat{A}_k^{(r)}=\sum_{t=k}^{R+1} 
       p_t^{(r)}\left( 
		      A_t^{(r)}-A_{t-1}^{(r)}
		\right) 
    \eeq

    \beq 
       A_j^{(r)}=\sum_{t=0}^{j}
       \frac{1}{p_t^{(r)}}\left( 
		      \widehat{A}_t^{(r)}-\widehat{A}_{t+1}^{(r)}
		\right) 
		\label{FT}
    \eeq
The RFT involved in this paper will always concern {\it cross-overlaps}, 
{\it i.e. lower indices}. In the case we are on a $4-$replica tree with two 
passive, direct overlaps, {\it e.g.} for $r<s$ we shall use $p_t^{(r,s)}$
(see sections 3.2, 4.1) with 

     \begin{eqnarray} 
	p_t^{(r,s)} = p_t & \, \, & t \leq r \nonumber \\ 
        p_t^{(r,s)} = 2p_t & \, \, & r < t \leq s \label{poids} \\
        p_t^{(r,s)} = 4p_t & \, \, & r < s < t \nonumber 
      \end{eqnarray}
and obvious extensions for $r=s$ or when more passive overlaps are involved.

\section{Kernels as RFT on a tree} 

Consider the $4-$replica matrix $M^{\alpha \beta; \gamma \delta}$ 
which we choose to parametrize as follows: 

\begin{itemize}
\item[(i)] on the {\it Replicon}-like configurations of the $4-$replica tree: 
      {\it i.e.} $\alpha \cap \beta \equiv \gamma \cap \delta=r$ 

         \beq
            M^{\alpha \beta; \gamma \delta} 
	    =
	    M^{r;r}_{u;v}
         \eeq
with the lower indices 

    \[
	\begin{array}{c}
              \max(\alpha \cap \gamma, \alpha \cap \delta)=u , \\
    	      \max(\beta \cap \gamma, \beta \cap \delta)=v   \\
	\end{array} 
	\hspace{1cm}
        u,v \geq r+1
     \]

\item[(ii)] The other configurations of the $4-$replica tree belong exclusively to the 
so called {\it Longitudinal-Anomalous} (LA) component

         \beq
            M^{\alpha \beta; \gamma \delta} 
	    =
	    M^{r;s}_{t}
	    \equiv
            {_A}M^{r;s}_{t}
         \eeq
where $\max(\alpha \cap \gamma, \alpha \cap \delta, \beta \cap \gamma, 
\beta \cap \delta)=t$, and where it may happen, accidentally, that $r=s$.
\end{itemize} 

The upper indices take values $0,1,\dots,R$ (for the special problem 
considered here, $R+1 \equiv (\alpha \cap \alpha)$ is excluded). 
Lower indices take values $0,1,\dots,R+1$. These two sets of 
variables will also be referred as direct overlaps and cross-overlaps, 
respectively.  

We now show, that the ``kernels'' defined in ref.\cite{Cyrano1, Temesvari}, 
in terms 
of which it was possible to invert the $4-$replica matrix 
$M^{\alpha \beta; \gamma \delta}$, are nothing but appropriate RFT's 
on {\it lower} indices (cross-overlaps). 

\subsection{The Replicon component $ {_R}M$}

It was recognized quite early\cite{Cyrano2, Cyrano3} that ${_R}M^{r;r}_{u;v}$
(or the corresponding component of the inverse {\it i.e.} of the 
propagator ${_R}G^{r;r}_{u;v}$)was obtained via some double transform 
of a more elementary object, the ``kernel'' $K^{r;r}_{k;l}$
(or $F^{r;r}_{k;l}$ for the propagator). The kernel, identified 
with the Replicon eigenvalue $\lambda(r;k,l)$ (see below), was indeed 
explicitly written as\cite{Kondor}

	    \[
              \mbox{kinetic term} + 
	      \lambda(r;k,l) \equiv 
              K^{r;r}_{k;l}=
	    \]
	    \beq
             =
	     \sum_{u=k}^{R+1} p_u 
	     \sum_{v=l}^{R+1} p_v 
	     \left(
                   M^{r;r}_{u;v} -
		   M^{r;r}_{u-1;v}-
		   M^{r;r}_{u;v-1} +
		   M^{r;r}_{u-1;v-1}
             \right), 
	     \,\,\,\,   k,l \geq r+1
	     \label{R_kernel}
	     \eeq
which we recognize now as a double RFT. Note that 

\begin{itemize} 
\item[(i)] we have written $M$ instead of ${_R}M$. We shall see below why 
this does not make any difference. 

\item[(ii)] we have not used the RFT on the tree (with passive $r$) since 
here $k,l \geq r+1$, rendering its use trivial (and pedantic). 

\end{itemize}

The direct relationship is then obtained by inverting the double RFT, 

	   \beq
	     _R M^{r;r}_{u;v}
	     =
	     \sum_{k=r+1}^{u} \frac{1}{p_k} 
	     \sum_{l=r+1}^{v} \frac{1}{p_l} 
	     \left(
                   K^{r;r}_{k;l} -
		   K^{r;r}_{k+1;l}-
		   K^{r;r}_{k;l+1} +
		   K^{r;r}_{k+1;l+1}
             \right)
	     \label{inverse}
	     \eeq
This was, unknowingly, the type of relationship between  
${_R}G^{r;r}_{u;v}$ and $F^{r;r}_{k;l}$ that first came out in the 
unravelling of the bare propagators (for the Parisi limit, $R=\infty$)
As already mentioned, analogous results have also been recently obtained, through the use of 
$p-$adic theory, by Parisi and Sourlas\cite{Parisi2}.

\subsection{The LA-component ${_A}M$ } 

With the above defined RFT on the tree (\ref{poids}) we can now write 

     \beq
        M^{r;s}_t
	=
	\sum_{k=0}^t  \frac{1}{p_k^{(r;s)}}
	\left[
	      K^{r;s}_k -K^{r;s}_{k+1} 
	\right] 
	\label{LA_RFT}
      \eeq

     \beq
        K^{r;s}_k
	=
	\sum_{t=k}^{R+1} p_t^{(r;s)}
	\left[
	      M^{r;s}_t -M^{r;s}_{t-1} 
	\right] 
	\label{LA_kernel}
      \eeq

Corresponding equations may be written  relating the 
propagator $G^{r;s}_t$ to its kernel, obtained by RFT, $F^{r;s}_k$. 
Equation (\ref{LA_kernel}) generates the LA kernel $K^{r;s}_k$, including 
the limiting value $r=s$. Conversely, knowing the kernel everywhere, we can 
generate the mass operator ${_A}M$ for {\it all} values of $r,s$. 

In the Replicon configurations of the $4-$replica tree we have two lower indices
${_A}M^{r;r}_{u;v}$ and therefore 
 
       \beq
         {_A}M^{r;r}_{u;v}
	 =
	 \sum_{k=0}^{\max(u,v)} 
	 \frac{1}{p_k^{(r,u,v)}}
	 \left(
	       K^{r;r}_k - K^{r;r}_{k+1} 
	 \right)	 
	 \label{LA_mass}
       \eeq
Note that $p_k^{(r,u,v)} \equiv p_k^{(r,\min(u,v))}$ in the interval bounded 
by $\max(u,v)$. From (\ref{LA_mass}), it follows that

   \beq
       {_A}M^{r;r}_{u;v}
       =
        {_A}M^{r;r}_{u}+ 
	{_A}M^{r;r}_{v} -
	{_A}M^{r;r}_{r}, 
	\,\,\,\,  u,v \geq r+1
	\label{LA_relation}
   \eeq
where $_AM^{r;r}_t$ is given by (27) taken at $r=s.$ 
This relationship explains why on the righthand side of (\ref{R_kernel})
 one may use, indifferently, $M$ or ${_R}M$: indeed ${_A}M^{r;r}_{u;v}$ 
where $u,v \geq r+1$, is shown in (\ref{LA_relation}) to depend upon a 
{\it single}\footnote{Note that if one wishes to attach a second {\it lower} 
index to $G^{r;s}_t$ ({\it e.g.} for the sake of faithfully representing 
a propagator by one pair of lines) it has to be $\min(r,s,t)$. } 
lower index at a time, it is thus projected out under a 
{\it double} RFT.

The equations given above for the kernels  (\ref{R_kernel},\ref{LA_kernel}) 
can be taken as {\it defining the kernels} once the primary functions $M$ are known 
({\it e.g.} by loop expansion). It is shown below that the same objects then 
block-diagonalize the eigenvalue equations (section 4) and the 
Dyson equation (section 5).

\section{Eigenvalue equations}

We now turn to the eigenvalue equations for the matrix 
$M^{\alpha \beta ; \gamma \delta}$. The three eigenvector classes, 
as introduced by de Almeida and Thouless\cite{deAlmeida}, write 
$f^{\alpha \beta}$, $f^{\alpha \beta; \mu}$, 
$f^{\alpha \beta; \mu \nu}$  for the L, A and R sectors respectively. 

\subsection{The L-sector}

    \beq
       \frac{1}{2} \sum_{\gamma \delta}      
       M^{\alpha \beta ; \gamma \delta}
       f^{\gamma \delta} 
       =
       \lambda_L f^{\alpha \beta} ,
     \eeq

here the sum is over all the configurations of the $4-$replica tree. Spelling it out we get 

      \beq
         \sum_{s=0}^{R} 
	 \left\{
                \delta_{r;s}^{\tiny \mbox{Kr}} 
	        \sum_{u=r+1}^{R+1} \delta_u
		\sum_{v=r+1}^{R+1} \delta_v
		M^{r;r}_{u;v} 
		+
		\frac{\delta_s}{2} 
		\sum_{t=0}^{R+1} 
		\delta_t^{(r,s)} M^{r;s}_t 
         \right\} f^s = \lambda f^r
	 \label{L_eigenvalue}
      \eeq
where the $\delta_{r;s}^{\tiny \mbox{Kr}}$ term is the contribution of the
Replicon-like configurations of the $4-$replica tree. 

In terms of RFT's, this writes

     \beq
        \sum_{s=0}^{R} 
	        \left\{
		       \delta_{r;s}^{\tiny \mbox{Kr}}
		       K^{r;r}_{r+1;r+1} + 
		       \frac{1}{2} K^{r;s}_0 \delta_s
		\right\} f^{s}=\lambda_L f^r
      \eeq
Here we have used 

     \beq
     	\delta_t^{(r;s)} \equiv  
	p_t^{(r;s)}-p_{t+1}^{(r;s)} 
     \eeq 
and $p_t^{(r;s)}$ as in (\ref{poids}). We have also used the fact that, 
for a sum carried over the {\it full} definition interval $(0,R+1)$  
or when in the R-sector over $(r+1,R+1)$, one has 

   \beq
      \sum_b^{R+1} \delta_t A_t 
      =
      \sum_{b}^{R+1} p_t(A_t -A_{t-1})=\widehat{A}_b
      \label{passage}
    \eeq
{\it i.e.} yielding the RFT at its lower bound  
$b=0$ or $b=r+1$ respectively. Hence in eq. (\ref{L_eigenvalue}) 
the sum over $\delta_u$, $\delta_v$ is a 
double RFT. Therefore it projects out ${_A}M^{r;r}_{u;v}$ and yields the RFT 
at its lower bound value $b=r+1$ for ${_R}M^{r;r}_{u;v}$. 
For the single $\delta_t^{(r,s)}$ sum, we recover the single RFT of $M^{r;s}_t$ at its 
lower bound $b=0$.

\subsection{The A-sector}

     \beq
       \frac{1}{2} \sum_{\gamma \delta}      
       M^{\alpha \beta ; \gamma \delta}
       f^{\gamma \delta; \mu} 
       =
       \lambda_A f^{\alpha \beta; \mu} 
     \eeq

The idea is now the following: just like before we obtained simple 
relationships by taking RFT over lower indices, {\it i.e.} 
cross-overlaps, we now take one more step and do RFT on the 
convolution product of cross-overlaps. More precisely, we have the 
concatenations 

   \[
      \begin{array}{cccc}
         {\buildrel \bullet \over \alpha } \beta; {\buildrel \bullet \over \gamma} \delta  & 
	 {\buildrel \bullet\bullet \over \gamma} \delta ; {\buildrel \bullet\bullet \over \mu} &
         \mbox{and} & 
         {\buildrel \bullet \over \alpha} \beta ; {\buildrel \bullet \over \mu} 
       \end{array}
   \]
or
   \[
      \begin{array}{cccc}
         \alpha {\buildrel \bullet \over \beta } ; {\buildrel \bullet \over \gamma} \delta  & 
	 {\buildrel \bullet\bullet \over \gamma} \delta ; {\buildrel \bullet\bullet \over \mu} &
         \mbox{and} & 
         \alpha {\buildrel \bullet \over \beta}; {\buildrel \bullet \over \mu} 
       \end{array}
   \]

and two other possibilities depending on the configurations of the 
$5-$replica tree. The replica pairs with alike dots are those whose 
(cross-)overlap is to be Replica Fourier transformed.
We sum first over the $5-$replicas, 
at $\alpha \cap \beta=r$, $\gamma \cap \delta=s$, fixed and passive, to 
extract the $k$ RFT component (as in going from (7) to (8) )
and then we sum sum over $s$ to obtain 

     \beq
        \sum_{s=0}^{R} 
	        \left\{
		       \delta_{r;s}^{\tiny \mbox{Kr}}
		       K^{r;r}_{k;r+1} + 
		       \frac{1}{4} K^{r;s}_k \delta_s^{(k-1)}
		\right\} \widehat{f}^{s}_k=\lambda_A \widehat{f}^r_k.
      \eeq

Note that

\begin{itemize}

\item[(i)] the summation over the concatened lower indices 
yields the product $K_k f_k$ (as in \ref{convolution}). 

\item[(ii)] in the Replicon subspace, there is a {\it second} lower index 
which is {\it freely} summed over its complete range $(r+1,R+1)$ hence yielding 
the $b=r+1$ component of $K^{r;r}$ (as in (\ref{passage}))

\item[(iii)] the $s$ upper index summation, at fixed (RFT)
cross-overlap $k$, comes out as $\delta_s^{(k-1)}.$

\item[(iv)] the ``monochromatic'' RFT $\widehat{f}_k^{r}$ corresponds to an 
eigevector (see (\ref{poids_bis}-\ref{FT}))

	   \beq
	      f^{r}_t =
              \left\{  \begin{array}{cc}
                         0                       &    t< k-1 \\[0.5cm]
                   -\frac{1}{p_{k-1}^{(r)}}\widehat{f}^r_k &    t=k-1  \\[0.5cm]
		     \left(\frac{1}{p_k^{(r)}} -\frac{1}{p_{k-1}^{(r)}}\right)
		     \widehat{f}^r_k                 & t>k-1 
		       \end{array}
              \right.
           \eeq
that is {\it null} for the cross-overlaps smaller than the {\it resolution} 
$k$, and {\it independent} of the {\it cross-overlap} when larger 
than $k$. 

\item[(v)] The L case identifies with $k=0$ and 

	    \beq
               f^r=\frac{1}{p_0} \widehat{f}^r_0
	    \eeq
\item[(vi)] The full multiplicity associated with a given 
 $\widehat{f}_k$ is 
 
 	\beq
           \mu(k)= n \bar{\delta}_k
	   =n\left(\frac{1}{p_k} - \frac{1}{p_{k-1}}\right) 
	\eeq
\end{itemize}

\subsection{The R-sector}

     \beq
       \frac{1}{2} \sum_{\gamma \delta}      
       M^{\alpha \beta ; \gamma \delta}
       f^{\gamma \delta; \mu \nu} 
       =
       \lambda_R f^{\alpha \beta; \mu \nu} 
       \hspace{1cm} 
       \alpha \cap \beta \equiv \gamma \cap \delta 
       \label{R_eigenvalue}
     \eeq

If $\mu \cap \nu \neq \alpha \cap \beta $, then there is a single 
(independent) cross-overlap, {\it i.e.} those eigenvectors belong 
to the LA subspace. We thus have necessarily 

   \beq
     \alpha \cap \beta \equiv \gamma \cap \delta \equiv \mu \cap \nu 
   \eeq
and we now have a {\it double} set of concatenations, {\it e.g.}

   \[
      \begin{array}{cccc}
         {\buildrel \bullet \over \alpha } \beta; {\buildrel \bullet \over \gamma} \delta  & 
	 {\buildrel \bullet\bullet \over \gamma} \delta ; {\buildrel \bullet\bullet \over \mu} \nu&
         \mbox{and} & 
         {\buildrel \bullet \over \alpha} \beta ; {\buildrel \bullet \over \mu} \nu 
       \end{array}
   \]

where one set is exhibited and the other is its complement. Altogether 
four double sets depending on configurations of the $6-$replica 
tree (restricted to its Replicon sector through (\ref{R_eigenvalue})). 
The double RFT 
associated (block)-diagonalizes the eigenvalue equations 
(\ref{R_eigenvalue}) (the blocks are here of dimension $1\times 1$, instead 
of $(R+1)\times (R+1)$ in the LA sector): 

   \beq
      K^{r;r}_{k;l}f^{r}_{kl}= \lambda_R f^{r}_{kl}
   \eeq

   \beq
     K^{r;r}_{k;l} \equiv \mbox{kinetic term} + \lambda(r;k,l), 
     \,\,\,\, k,l \geq r+1  
   \eeq
The associated multiplicities can be written in term of $\bar{\delta}_{k}$ 
the (relative) multiplicity in pseudo-momentum or resolution space, with the proviso that, 
at the lower bound $b$ of the interval of definition, we have  

    \beq
       \bar{\delta}_b=\frac{1}{p_b} 
    \eeq

$b=0$ as for (\ref{mult}), for the LA sector, and $b=r+1$ for the R sector of 
direct overlap $r$.  Again, using $p_k^{(r)}$ instead of $p_k$ would amount to 
some trivial change in the $\bar{\delta}_{k}$ since $k,l \geq r+1$. 
One finds 

   \beq
      \mu(r;k,l)= \frac{n}{2}
      \bar{\delta}_k 
      \bar{\delta}_l \delta_r{(k,l)}
   \eeq
where $\delta_r{(k,l)} = p_r - (\alpha + 1)p_{r+1}$ and $\alpha = 0,1,2$
is the occupation of boxes $p_{r+1}$ by the cross-overlaps $k,l.$
For further use, this is conveniently separated as 

   \beq
      \mu \equiv \mu_{\rm reg} + \mu_{\rm sing} 
   \eeq

   \beq
      \mu_{\rm reg}(r;k,l)= \frac{n}{2}
      \bar{\delta}_k 
      \bar{\delta}_l \delta_r
   \eeq

   \beq
      \mu_{\rm sing}(r;k,l)
      =\left\{
              \begin{array}{cc} 
		  0                           &    k,l > r+1 \\[0.5cm] 
               -\frac{n}{2}\bar{\delta}_l   & k=r+1,\,  l > r+1 \\[0.7cm]
               \displaystyle{-n\sum_{s=0}^{r+1} \bar{\delta}_s} & k,l=r+1 
              \end{array} 
        \right.
	\label{mu_sing}
    \eeq
The total degeneracy becomes

    \beq
       \sum_{r=0}^R
       \sum_{k=r+1}^{R+1} 	
       \sum_{l=r+1}^{R+1} 
       \mu_{\rm reg}(r;k,l) 
       =
       \frac{n(n-1)}{2} 
    \]			
    \[
       \sum_{r=0}^R
       \sum_{k=r+1}^{R+1} 	
       \sum_{l=r+1}^{R+1} 
       \mu_{\rm sing}(r;k,l) 
       =
       -n(R+1) 
    \eeq			
and for the LA sector, from the $(R+1)\times (R+1)$ blocks 

    \beq
       (R+1)\sum_{k=0}^{R+1} \mu(k)= + n(R+1)
    \eeq
yielding back the appropriate count.

\section{Block-diagonalization of the  Dyson equation}

Whatever has been found above for the matrix $M$ and the associated kernels 
$K$ ($K^{r;r}_{k;l}$ in the R-sector, $K^{r;s}_k$ in the LA one) can be 
repeated for the inverse matrix $G$ (the propagator matrix) and the 
associated kernels $F$. Expresssing their relationship via 

     \beq 
        \sum_{\gamma \delta} 
        M^{\alpha \beta; \gamma \delta} 
	G^{\gamma \delta; \mu \nu}
        = 
	\delta_{\alpha \beta; \mu \nu}^{\tiny \mbox{Kr}} 
     \eeq
one is now able to block-diagonalize this matrix equation. To do this, one uses

\begin{itemize} 

\item[(i)] a double RFT in the Replicon sector yielding 

	   \beq
              K^{r;r}_{k;l} F^{r;r}_{k;l} = 1
	      \label{R_Dyson}
	   \eeq
a result which can also be obtained {\it via} $p-$adic theory according 
to Parisi and Sourlas\cite{Parisi2}. 

\item[(ii)] a single RFT in the LA sector, {\it viz} 

	    \beq
               \sum_{t=0}^R
	       \left(
	             \delta_{r;t}^{\tiny \mbox{Kr}}
		     \Lambda_k(r) 
		     +
		     \frac{1}{4}K^{r;t}_k 
		     \delta_t^{(k-1)}
	        \right) 
		\left(
	             \delta_{t;s}^{\tiny \mbox{Kr}}
		     \frac{1}{\Lambda_k(t)} 
		     +
		     \frac{1}{4}F^{t;s}_k 
		     \delta_s^{(k-1)}
	        \right)= \delta_{r;s}^{\tiny \mbox{Kr}}
		\label{LA_Dyson}
             \eeq
\end{itemize}
Here 

  \[
     \Lambda_k(r)
     =\left\{
	     \begin{array}{ccc} 
	        K^{r;r}_{r+1;r+1} & \mbox{if} & k \leq r+1 \\[0.5cm]	
		K^{r;r}_{r+1;k}   & \mbox{if} & k >    r+1 \\[0.5cm]
	     \end{array}
      \right.
   \]
and in (\ref{LA_Dyson}) we have used (\ref{R_Dyson}) to obtain 
$1/\Lambda_k(r)$.

After division by $\delta_s^{(k-1)}/4$ one gets the Dyson's equation, 
relating the kernels of inverses 

 \beq
    F^{r;s}_k
    =
    -\frac{1}{\Lambda_k(r)} K^{r;s}_k \frac{1}{\Lambda_k(s)} - 
     \sum_{t=0}^R  
     \frac{1}{\Lambda_k(r)} K^{r;t}_k 
     \frac{\delta_t^{(k-1)} }{4} F^{t;s}_k
     \label{LA_Dysonbis} 
  \eeq
or 

 \beq
    \overline{F}^{r;s}_k
    =
    K^{r;s}_k  
    -\sum_{t=0}^R  
     K^{r;t}_k 
     \frac{\delta_t^{(k-1)} }{4 \Lambda_k(t)} 
     \overline{F}^{t;s}_k 
  \eeq
if 

   \beq
       \overline{F}^{r;s}_k \equiv - \Lambda_k(r) F^{r;s}_k \Lambda_k(s)
   \eeq

Note that if one writes out explicitly $\mbox{tr}MG$, one has to sum 
over multiplicities, and an {\it exact} cancellation occurs then between 
the terms in 
    
    \[
      \mu_{\rm sing}(r;k,l)K^{r;r}_{k;l} F^{r;r}_{k;l}
    \] 
coming from the  R-sector, and the terms 

    \[
      \mu(k)K^{r;r}_{k;r+1} F^{r;r}_{k;r+1} 
   \]
originating from the LA-sector. 

\section{Conclusion} 

Assuming that via perturbation expansion we have computed $M$ up to 
some loop order then, to obtain the corresponding $G$ one would do the 
following:

\begin{itemize}

\item[(i)] compute the kernels associated with $M$: 
           $K^{r;r}_{k;l}$ via (\ref{R_kernel})  and $K^{r;s}_k$ via 
           (\ref{LA_kernel})

\item[(ii)] obtain the corresponding kernel associated with $G$, the inverse 
            of $M$, {\it viz} $F^{r;r}_{k;l}=1/K^{r;r}_{k;l}$ via 
	    (\ref{R_Dyson}) and $F^{r;s}_k$ via a solution of 
	    (\ref{LA_Dysonbis}). This last step has been shown to be analytically 
	    feasible\cite{Cyrano1} if, as it turns out to be at
	    the zero loop level, 
                 $ K^{r;s}_k=K_k(\min r,s) $
	      
\item[(iii)] knowing $F^{rr}_{k;l}$ one obtains ${_R}G^{r;r}_{u;v}$ via  
            (\ref{inverse}, transposed for $F,G$). The knowledge
            of $F^{r;s}_k$ gives ${_A}G^{r;r}_{u;v}$, $u,v \geq r+1$, 
	    via (\ref{LA_mass}, transposed for $F,G$) and the other 
	    components of ${_A}G^{r;s}_t$ via 
	    (\ref{LA_RFT}, transposed for $F,G$). 

\end{itemize} 

The sum over multiplicities in ``resolution'' space generated by  the 
inverse RFT directly constructs $\mu_{\rm reg}(r;k,l)$. The cancellations described just above,  
that occur between R contributions with $\mu_{\rm sing}$ weight and 
the (Replicon-like) contributions arising as diagonal terms in the 
block-diagonalized LA sector are already taken care of since, via the RFT on the 
tree, we directly (block) diagonalize the eigenvalue equation and the Dyson
equation.  

To conclude, let us emphasize that the 
``conservation law''\footnote{On the ultrametric tree! {\it I.e.} resulting 
in the pseudo-momentum equality for a $2-$point function, but 
ultrametric inequalities for a $p-$point function, $p \geq 3$.} on
pseudo-momentum ($k$, or $k$ and $\ell$ in the double RTF) which
allows for the ``mass-operator" diagonalization (just like the
ordinary Fourier Transform under translational invariance)
should play a central role in the derivation of the much
wanted ``Feynman Rules" for the field theory of the spin glass.

\vspace{1.5cm}
The authors are grateful to G.Parisi and N.Sourlas for making 
available their unpublished results.

\end{document}